\documentclass[a4paper,fleqn,usenatbib]{mnras}

\usepackage[T1]{fontenc}
\usepackage{ae,aecompl}

\usepackage{graphicx}	% Including figure files
\usepackage{amsmath}	% Advanced maths commands
\usepackage{amssymb}	% Extra maths symbols
\usepackage{breakurl}

\usepackage{txfonts} %arxiv fonts

\newcommand{\etal}{{et al}\/.}

\title[Spectral ageing in the era of big data]{Spectral ageing in the era of big data: integrated vs resolved models}

\author[J.J.~Harwood \etal]{Jeremy J.\ Harwood\thanks{E-mail: jeremy.harwood@physics.org}
\\ASTRON, The Netherlands Institute for Radio Astronomy, Postbus 2, 7990 AA, Dwingeloo, The Netherlands
}

\date{Accepted XXX. Received YYY; in original form ZZZ}

\pubyear{2016}

\begin{document}
\label{firstpage}
\pagerange{\pageref{firstpage}--\pageref{lastpage}}
\maketitle

\graphicspath{{./images/}}

\begin{abstract}

Continuous injection models of spectral ageing have long been used to determine the age of radio galaxies from their integrated spectrum; however, many questions about their reliability remain unanswered. With various large area surveys imminent (e.g. LOFAR, MeerKAT, MWA) and planning for the next generation of radio interferometer well underway (e.g. ngVLA, SKA), investigations of radio galaxy physics are set to shift away from studies of individual sources to the population as a whole. Determining if and how integrated models of spectral ageing can be applied in the era of big data is therefore crucial. In this paper, I compare classical integrated models of spectral ageing to recent well resolved studies that use modern analysis techniques on small spatial scales to determine their robustness and validity as a source selection method. I find that integrated models are unable to recover key parameters and, even when known a priori, provide a poor, frequency dependent description of a source's spectrum. I show a disparity of up to a factor of 6 in age between the integrated and resolved methods but suggest, even with these inconsistencies, such models still provide a potential method of candidate selection in the search for remnant radio galaxies and in providing a cleaner selection of high redshift radio galaxies in $z - \alpha$ selected samples.

\end{abstract}


\section{Introduction}
\label{intro}

Radio galaxies, while only comprising a small fraction of the total population, are widely believed to play a key role in the formation and evolution of the galactic population due to their ability to provide the additional energy required to affect star formation rates in a self-regulating manner through the expulsion of cold gas and heating of the inter-galactic/cluster medium (AGN feedback; \citealp{croton06, bower06}). While great strides have been made in understanding the underlying physics of these galaxies, studies have remained limited to a small number of bright, nearby, objects meaning that the applicability of such findings to the wider population has remained limited.

Large area surveys such as the LOw Frequency ARray (LOFAR) Two-metre Sky Survey (LOTSS; \citealp{shimwell16}), the Murchison Widefield Array (MWA) GLEAM survey \citep{wayth15}, and the MeerKAT MIGHTEE survey \citep{jarvis12}, combined with the vast data volumes and sample sizes that will be provided by future telescopes such as the next generation VLA (ngVLA) and the Square Kilometer Array (SKA) mean that in the upcoming decades, one will be able to investigate the radio galaxy population in much greater detail.

One of the key outstanding issues is that of the characteristic age, dynamics and life-cycle of the powerful \citet{fanaroff74} class I (FR I) and II (FR II) radio galaxy populations. As the lobes and plumes which create the distinctive morphology of FR Is and IIs emit via the synchrotron process, one of the most commonly used methods in determining the time since a source first became active is through models of spectral ageing. These models describe the preferential cooling of higher energy electrons as a function of time due to synchrotron and inverse-Compton losses, resulting in a more highly curved spectrum in older regions of plasma (e.g. the JP and KP models; \citealp{kardashev62, pacholczyk70, jaffe73}) from which the total age of the source can be determined. However, many outstanding questions remain about the robustness and reliability of these models in determining the intrinsic age of a source.

\begin{table}
\centering
\caption{Source properties}
\label{targets}
\begin{tabular}{llcccc}
\hline
\hline
Name&IAU Name&$z$&178 MHz&Type&Ref.\\
&&&Flux (Jy)&&\\
\hline
3C452&J2243$+$394&0.081&130&II&1, 2\\
3C223&J0936$+$361&0.137&9.0&II&1, 2\\
3C438&J2153$+$377&0.290&48.7&II&3\\
3C28&J0053$+$261&0.195&17.8&II (R)&3\\
3C31&J0104$+$321&0.017&18.3&I&4\\
NGC3801&J1137$+$180&0.011&3.4&I&5\\
\hline
\end{tabular}
\vskip 5pt
\begin{minipage}{8.2cm}
Summary of sources used in the spectral age model fitting. `Type' refers to the \citet{fanaroff74} morphological classification where (R) indicates a remnant galaxy in which the central engine is no longer active. `$z$' list the redshift and `178 MHz Flux' the integrated flux of each source at 178 MHz. `Ref' column gives the reference for the origin of the values, denoted as follows: (1) \citet{harwood16}; (2) \citet{harwood17a}; (3) \citet{harwood15}; (4) \citet{heesen14}; (5) \citet{heesen16}.
\end{minipage}

\end{table}

While many of the issues surrounding these models have been addressed when considering radio galaxies on well resolved scales (e.g. \citealp{harwood13, heesen14, harwood15, harwood17a}) it is unclear whether models which consider only the integrated flux of a source can replicate these results. The application of these models to large samples of galaxies is mainly limited by two factors; computational speed and the availability of suitable ancillary data. While fitting of models on well resolved scales allows us to probe more deeply the dynamics and underlying physics which drive emission from these sources, the computational resources, ancillary data, and human interaction required for highly accurate image alignment required to perform fitting on such small scales is unlikely to be feasible on reasonable timescales for a large number of sources. It is therefore crucial that we understand whether current models which consider only the integrated flux (where resource demands are much lower) are able to provide accurate spectral ages and a robust source selection methods based on their current phase of activity (e.g. active or remnant\footnote{I preference the term remnant over relic to avoid confusion with the similarly named cluster relics.}).

The most commonly used models for such purposes are the Continuous Injection (CI) model proposed by \citet{jaffe73} for active sources and the CI off (also known as the KGJP) model presented by \citet{komissarov94} for remnant radio galaxies in which the central engine is no longer active. Both of these variants are based on synchrotron and inverse-Compton losses described by the standard JP model (see \citealp{harwood13, harwood15} for a detailed analysis of these models and \citealp{longair11} for a full derivation) but assume an injection of fresh, zero-age plasma either for the lifetime of the source (CI) or some fraction of its age after which no additional injection of plasma occurs (CI off). This, in theory, allows the age of a radio galaxy to be determined from the spectrum of the integrated flux alone, where the fresh plasma is provided by the acceleration regions and the aged plasma by emission from the source's lobes/plumes.

With multiple large area surveys currently underway and the science possible with various instrument configurations for the next generation of radio telescope currently being discussed, it is crucial for maximising the potential scientific output of these projects that it is determined which analytical and selection methods can be used reliably. In this paper, I therefore investigate the robustness of CI models compared to well resolved studies recently undertaken that are constrained by broad bandwidth observations and modelling on small spatial scales and discuss the impact and implications of these results on the future of radio galaxy science.

\subsection{Outstanding questions addressed in this paper}
\label{questions}

Given the importance of such models in both the study of radio galaxy populations and in the planning of the science possible with the next generation of radio telescopes, in this paper I will address 2 primary questions:

\begin{enumerate}
\item When parameters are well constrained by resolved studies, do CI and CI off models provide accurate spectral ages?\\
\item Are integrated models able to reliably distinguish between the different phases of a radio galaxy's life cycle?
\end{enumerate}

In Section \ref{method} I give details of target selection and the analysis undertaken and in Section \ref{discussion} I present the model fitting results and discuss these findings in the context of the aims outlined above. Throughout this paper, a concordance model in which $H_0=71$ km s$^{-1}$ Mpc$^{-1}$, $\Omega _m =0.27$ and $\Omega _\Lambda =0.73$ is used \citep{spergel03}.

\begin{figure*}
\centering
\includegraphics[angle=0,width=17.8cm]{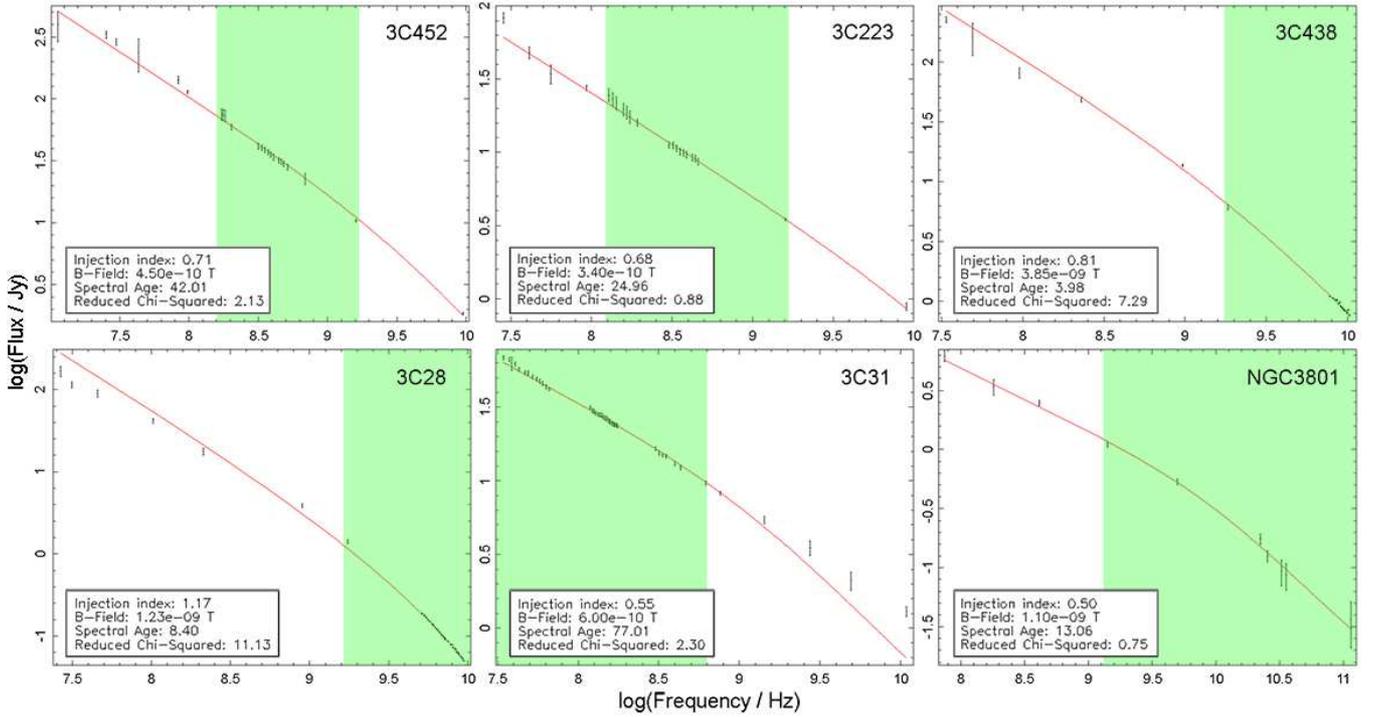}
\caption{Plots of flux against frequency with the best fitting CI model overlaid (redline). Note that 3C28 is fitted with the CI off model due to the remnant nature of the source. Model parameters, ages (in Myr) and statistics are shown in bottom left corner of each panel. Shaded regions indicate the data used for the resolved fitting and the limited frequency coverage case.}
\label{plots}
\end{figure*}

\section{Target Sources and Spectral Analysis}
\label{method}

\subsection{Source sample and archival data}
\label{datareduction}

In order to address the aims outlined in Section \ref{questions}, the six radio galaxies for which spectral age fitting has previously been performed on small spatial scales using broad bandwidth observations and the methods of \citet{harwood13, harwood15} were selected. This not only provides a uniform methodology across our sample (summarised in Section \ref{spectralanalysis}), but also reduces the effects of many of the known problems that affect historical studies, for example cross-lobe age variations, and provides tight constraints on crucial models parameters such as the spectrum produced by the initial electron energy distribution (the `injection index'\footnote{I define the spectral and injection index such that $S \propto \nu^{-\alpha}$}) and magnetic field strength. The sample comprises of two FR Is, three FR IIs, and one remnant FR II in which the central engine is no longer active. Reduction of the observations for each source are detailed in the corresponding original study which, along with a summary of target properties, are listed in Table \ref{targets}.

While the data required for resolved studies are limited to observations matched in terms of resolution, a major benefit of integrated fitting is the ability to expand the frequency coverage by using low resolution archival data. This is likely to be common practice for the currently ongoing wide area surveys and so additional low frequency archival flux values were also obtained from the literature. For the three FR II sources (including the remnant source, 3C28) these were taken from the study of 3C sources at low frequencies by \citet{laing80} providing data points at 22.25, 26.3, 38.0, 86.0, 178 and 750 MHz. For the FR I source NGC 3801, values at 74.0 MHz from the study of \citet{cohen07} and at 178 and 408 MHz from the Parkes catalogue (PKSCAT90, \citealp{wright90}) were obtained. As the recent study of 3C31 by \citet{heesen16} already includes frequencies between 33.9 MHz and 10.7 GHz, no additional archival data was required in this case but note that only frequencies between 52 and 609 MHz were used during the resolved fitting.

\subsection{Spectral analysis}
\label{spectralanalysis}

\subsubsection{Resolved model fitting}
\label{resolvedmethod}

As noted above, spatially resolved spectral age fitting has previously been preformed for all sources in the sample using the Broadband Radio Astronomy ToolS (\textsc{brats}\footnote{http://www.askanastronomer.co.uk/brats}) software package. A detailed description of the software's usage and methodology has previously been carried out by \citet{harwood13, harwood15}, the accompanying cookbook, and the respective authors for each source (Table \ref{targets}) and so I do not duplicate such a discussion here, but for completeness provide a brief summary of the implementation and fitting procedure used.

For each object in the sample, a region loosely encompassing the source was defined along with a background region well away from any strong emission in order to determine the off-source thermal noise. The radio images were then loaded in to \textsc{brats} and an initial source detection performed based on the RMS noise value and a 5$\sigma$ cut-off. The `adaptiveregions' command was then used to automatically define small, pixel size regions in order to reduce the effects of the superposition of spectra as discussed by \citet{harwood13, harwood15}. Fitting of the numerically calculated model spectrum to the observational data was then performed using a grid search over the potential age space (defined by each study individually) and a golden ratio search for the normalisation, with the best fitting model determined through $\chi^{2}$ minimisation. Images of age as a function of position and statistical values were then exported for further analysis and to determine the robustness of the fitting results.

\begin{table*}
\centering
\caption{Summary of model fitting results (full frequency coverage)}
\label{fullfittingresults}
\begin{tabular}{lccccccccccccc}
\hline
\hline
Name	&	$\chi^{2}_{Red}$	&	Rejected	&	Confidence	&	$T_{Int}$	&	+	&	-	&	$T_{Res}$	&	+	&	-	&	$T_{Diff}$	&	+	&	-	\\
	&		&		&	($\%$)	&	(Myr)	&		&		&	(Myr)	&		&		&	(Myr)	&		&		\\
\hline

3C452	&	2.13	&	Yes	&	$>99$	&	42.01	&	1.31	&	1.50	&	89.05	&	8.56	&	7.14	&	-47.04	&	8.66	&	7.30	\\
3C223	&	0.88	&	No	&	$<68$	&	24.96	&	2.51	&	3.55	&	77.95	&	13.42	&	11.73	&	-52.99	&	13.65	&	12.26	\\
3C438	&	7.29	&	Yes	&	$>99$	&	3.98	&	0.02	&	0.04	&	3.00	&	0.05	&	0.08	&	0.98	&	0.05	&	0.09	\\
3C28	&	8.40	&	Yes	&	$>99$	&	8.40	&	0.04	&	0.03	&	12.42	&	0.44	&	0.44	&	-4.02	&	0.44	&	0.44	\\
3C31	&	2.30	&	Yes	&	$>99$	&	77.01	&	3.49	&	2.68	&	169.99	&	32.72	&	30.15	&	-92.98	&	32.91	&	30.27	\\
NGC3801	&	0.69	&	No	&	$<68$	&	12.06	&	1.50	&	1.12	&	2.00	&	0.40	&	0.40	&	10.06	&	1.55	&	1.19	\\

\hline
\end{tabular}

\vskip 5pt
\begin{minipage}{17.8cm}
Results of the CI model fitting in the full frequency coverage case. `$\chi^{2}_{Red}$' lists the reduced $\chi^{2}$ values for the CI model fitting with the `Rejected' column stating if the model can be statistically rejected at the level given in the `Confidence' column'. `$T_{Int}$' lists the age of the source using the integrated flux and CI model, `$T_{Res}$' the age using resolved methods and the single injection JP model, and  `$T_{Diff}$' the difference between the two ages where $T_{Diff} = T_{Int} - T_{Res}$.

\end{minipage}
\end{table*}

For many of these studies, multiple model types were fitted to the data such as the KP model \citep{kardashev62, pacholczyk70} and the more recent Tribble model \citep{tribble93,hardcastle13a,harwood13}. However, I limit this study to a comparison of the JP model, with the alternatives generally being considered either unphysical (KP) or with the integrated version being indistinguishable from the JP model for the given observational constraints (Tribble). The integrated counterpart to the JP model (described below) is therefore almost exclusively used in the fitting of unresolved sources and in surveys. In the resolved case where such distinctions can (and have) be made, variations between models with respect to their spectral age and goodness-of-fit are known to be small relative to the results of Section \ref{discussion} and so can be safely ignored in the current context.

\subsubsection{Integrated model fitting}

As described in Section \ref{intro}, the standard (JP) CI model of spectral ageing provides the spectrum produced by a combination of continuous injection of freshly accelerated plasma and regions that have been subject to synchrotron and inverse-Compton losses described by the JP model \citep{kardashev62, jaffe73}. This model was further expanded by \citet{komissarov94} to produce the CI off model which allows freshly accelerated plasma to be injected for only a fraction of the source's total age, describing the spectrum of current inactive sources (e.g. radio galaxy remnants). Both of these models have been discussed previously in great detail and so I refer the interested reader to the original and subsequent studies (e.g. \citealp{kellermann64, carilli91, murgia99, brienza16}), again providing here only a summary of the models and their implementation.

Assuming an electron population radiating in a fixed magnetic field that is subject to synchrotron and inverse-Compton losses such that \begin{equation}\label{poploss} N(E,\theta,t) = N_{0} E^{-\delta} (1 - E_{T} E)^{-\delta-2}\end{equation} where $E_{T}$ are the model dependent losses (in this case JP losses) which are a function of the pitch angle of the electrons to the magnetic field, $\theta$, and the time since they were accelerated, $t$, and in which a fresh supply of plasma is continuously injected with an energy distribution described by \begin{equation}\label{initialdist} N(E) = N_0E^{-\delta} \end{equation} the integrated spectrum results in a broken power law (the CI model). At energies below a break frequency of \begin{equation}\label{break} \nu_{b} = \frac{9 \, c_{7} \, B} {4 t^{2} (B^{2} + B_{CMB}^{2})}\end{equation} where $t$ is the age of the source in Myr, B is the magnetic field strength in nT, $B_{CMB} = 0.318(1 + z)^{2}$ nT is the equivalent magnetic field strength for the CMB and $c_{7} = 1.12 × 10^{3}$ nT$^{3}$ Myr$^{2}$ GHz is a constant defined by \citet{pacholczyk70}, the spectral index is, in theory, given by \begin{equation}\label{alphalow} \alpha_{low} = \frac{\delta-1}{2} \end{equation} and above this break frequency \begin{equation}\label{alphahigh} \alpha_{high} = \alpha_{low} + 0.5\end{equation} For inactive sources \citet{komissarov94} show that in addition to this break, at frequencies above \begin{equation}\label{breakoff} \nu_{b,off} = \frac{(t_{on}+t_{off})^{2}\,\nu_{b}}{t_{off}^{2}} \end{equation} where $t_{on}$ is the time in Myr for which the source was active and $t_{off}$ the time since the central engine switched off, that the spectrum drops off exponentially (the CI off model).

\begin{table*}
\centering
\caption{Summary of model fitting results (limited frequency coverage)}
\label{resfittingresults}
\begin{tabular}{lcccccccccccc}
\hline
\hline
Name	&	$\chi^{2}_{Red}$	&	Rejected	&	Confidence	&	$T_{Int}$	&	+	&	-	&	$T_{Res}$	&	+	&	-	&	$T_{Diff}$	&	+	&	-	\\
	&		&		&	($\%$)	&	(Myr)	&		&		&	(Myr)	&		&		&	(Myr)	&		&		\\
\hline

3C452	&	0.23	&	No	&	$<68$	&	52.05	&	4.04	&	3.93	&	89.05	&	8.56	&	7.14	&	-37.00	&	9.47	&	8.15	\\
3C223	&	0.43	&	No	&	$<68$	&	25.99	&	6.78	&	6.71	&	77.95	&	13.42	&	11.73	&	-51.96	&	15.04	&	13.51	\\
3C438	&	8.34	&	Yes	&	$>99$	&	81.04	&	7.11	&	5.87	&	3.00	&	0.05	&	0.08	&	78.04	&	7.11	&	5.87	\\
3C28	&	2.01	&	Yes	&	$>99$	&	29.40	&	0.94	&	2.32	&	12.42	&	0.44	&	0.44	&	16.98	&	1.04	&	2.36	\\
3C31	&	0.12	&	No	&	$<68$	&	84.99	&	6.40	&	7.43	&	169.99	&	32.72	&	30.15	&	-85.00	&	33.34	&	31.05	\\
NGC3801	&	0.75	&	No	&	$<68$	&	13.06	&	1.42	&	0.89	&	2.00	&	0.40	&	0.40	&	11.06	&	1.48	&	0.98	\\

\hline
\end{tabular}
\vskip 5pt
\begin{minipage}{17.8cm}
Results of the CI model fitting in the limited frequency coverage case.`$\chi^{2}_{Red}$' lists the reduced $\chi^{2}$ values for the CI model fitting with the `Rejected' column stating if the model can be statistically rejected at the level given in the `Confidence' column'. `$T_{Int}$' lists the age of the source using the integrated flux and CI model, `$T_{Res}$' the age using resolved methods and the single injection JP model, and  `$T_{Diff}$' the difference between the two ages where $T_{Diff} = T_{Int} - T_{Res}$.
\end{minipage}
\end{table*}

Analytical solutions are possible for both of these models providing a significant reduction in computational overheads compared to their resolved counterpart which must be calculated numerically. As for $t_{off} = 0$ the CI off model tends to the standard CI model spectrum, \textsc{brats} uses the analytical solutions of \citet{komissarov94} to calculate the model spectra in both cases. In the fitting of these models to observations, \textsc{brats} uses an identical method to that implemented for the resolved model fitting (Section \ref{resolvedmethod}), exploring the parameter space using a golden ratio search for the normalisation and a grid search to determine the active source time (where for the CI model $t_{on}=t$). In the case of the CI off model, a grid search is also performed for $t_{off}$ resulting in an additional degree of freedom.

Applying the methods described above, \textsc{brats} was used to fit the CI model to the 5 active sources and the CI off model to the remnant source 3C28. In order to perform a preliminary test on the frequency dependence of the integrated models, fitting was run using both the full frequency coverage (including additional archival data) and limited to that of the original studies to replicate the impact on differing analysis scenarios. As, in the context of radio galaxies, the CI models can be thought of as the summation of the resolved JP spectra under a set of assumed conditions, the resolved studies should provide better constraints with respect to the model parameters. The injection index was therefore fixed to that derived in the original papers along with the estimates of the magnetic field strength. This assumption and its impact on large sample studies is discussed further in Section \ref{discussion}.

\vspace{-2.5mm}
\section{Results and Discussion}
\label{discussion}

\subsection{Spectral age and model fitting}
\label{age}

\subsubsection{Full frequency coverage}
\label{full}

From the numerical results given in Table \ref{fullfittingresults} and the corresponding plots shown in Fig. \ref{plots}, it is immediately evident that a disparity exists both in terms of the goodness-of-fit and the spectral ages between those derived from an integrated and a well resolved standpoint. Of the six sources tested, over half can be statistically rejected at the 99 per cent confidence level when the full frequency coverage is considered which is in stark contrast to the resolved fitting where in only one case (3C438) is the model statistically rejected, most likely due to the mixing of electron populations \citep{harwood15}.

Of the remaining rejected model fits, 3C452 is due to a discrepancy in the very low frequency spectrum. From Fig. \ref{plots} one can see that at frequencies greater than $116$ MHz the spectrum is well fitted by the CI model, a feature which has been previously noted by \citeauthor{harwood16} (\citeyear{harwood16, harwood17a}). A better goodness-of-fit can be obtained by steepening the injection index  but this only serves to further widen the disparity in the derived ages and can therefore be considered an upper limit on the CI model's age. 

Similarly, 3C31 is also well fitted over part of its frequency range but in this case it is the higher frequencies ($> 1$ GHz) where the divergence between the model and observations occur. In order to be well fitted, the curvature at lower frequencies must be ignored and the break in the spectrum must occur at frequencies $> 10$ GHz; however, resolved fitting, classical dynamical models, and alternative models of the FR I spectrum such as advection models \citep{heesen16} suggest an age in excess of $150$ Myr. From Equation \ref{break} one can see that this break must occur below $328$ MHz but, while such a break is seen in the observations, it does not conform to the change of $\Delta \alpha = 0.5$ expected by the CI model (Equation \ref{alphahigh}) and so the flux at high frequencies is greater than expected.

The spectrum of 3C28, for which the CI off model was fitted, proves to be an interesting case. The model is statistically rejected due to a flatter than expected spectrum at low frequencies and a sharper than expected cutoff at higher energies. Such behaviour has previously been observed by \citet{brienza16} in the recently discovered BLOB1 remnant galaxy suggesting that this may be a common feature of the integrated spectrum of remnant sources. It was noted by \citet{harwood15} that the injection index for 3C28 may represent an upper limit due to additional processes unaccounted for in the standard (resolved) ageing models; however, even when fitting for a more standard value for the injection index of $0.7$ the model, although better fitted at low frequencies, is still rejected at the 99 per cent confidence level and unable to account for the rapid steepening at shorter wavelengths. Variations in the low frequency spectrum of all sources and the observed rapid steepening of high frequency spectrum in remnant galaxies is discussed in the context of source selection further in Section \ref{selection}.

\vspace{-2mm}
\subsubsection{Limited frequency coverage}
\label{limited}

While it is clear that even when the best frequency coverage available is considered many problems exist in the fitting of integrated models of spectral ageing, the frequency limited sample also highlights additional dangers in applying these models to surveys and large samples. From Table \ref{resfittingresults} one can see that two galaxies, 3C452 and 3C31, are no longer statistically rejected and would therefore be interpreted as robust over such a frequency range. As was mentioned in Section \ref{full}, both of these sources are well fitted by the models over part of their spectrum and it is this region of frequency space which is being considered when only the resolved observations are used. Of the remaining sources, 3C438 and 3C28 remain rejected but similar arguments can be made simply by shifting the frequency range to a region where the model is well fitted (e.g. at $>\,$4 GHz). The goodness-of-fit is therefore highly dependent on our choice of frequency coverage if the spectrum is not constrained well above, below, and in the region of the spectral break.

For sources such as those considered here this is unlikely to be an issue in the majority of cases due to large quantities of data being available for these bright, nearby objects. However, future instruments (particularly those in the southern hemisphere) which will survey large areas of sky at much higher sensitivities will initially not have the luxury of archival measurements spanning such a wide frequency range. The majority of any sample taken from these surveys will therefore be strongly impacted by such effects and emphasises the need for a wide frequency coverage when fitting models to the integrated flux.

\vspace{-3mm}
\subsubsection{Evaluation and recommendations}
\label{cause}

The rejection of the majority of the integrated models, unsurprisingly, leads to large discrepancies in the age of the source; however, even in the cases where the models do provide a good description of the observed spectrum, a large disparity remains ranging between an underestimation of the ages by factor of $\approx 3$ for 3C223 and an overestimation of the age by a factor of $\approx 6$ for NGC 3801. Considering the dynamics of these sources, this would place the characteristic advance speed of 3C223 at approximately 5 per cent the speed of light, much faster than the commonly observed values of 1 to 3 percent for similar sources \citep{myers85, alexander87, liu92, harwood13, harwood15, harwood17a}.

While the exact cause of the discrepancy between the integrated and resolved spectrum of radio galaxies is not immediately clear, as the CI models are based on the single injection JP model it must at least in part be due to a breakdown in the assumptions made when integrating the individual spectra over the entire source.

One such assumption is that parameters such as magnetic field strength, external environment, and the expansion of the lobes/plumes are constant over the course of the galaxy's active lifetime. A possible solution therefore lies in an extension of that discussed by \citet{harwood16} in the context of determining the injection index, who suggest that for FR IIs the integrated spectrum may differ due to it being a sum of the history of the source where conditions are non-uniform over its lifetime. If, in addition to synchrotron and inverse-Compton losses, radio galaxies are also subject to time-dependent adiabatic losses (e.g. \citealp{murgia99}) and/or a mixing of electron populations, simulations suggest that the observed integrated spectrum can be driven to steeper values \citep{kapinska15}. Such time-dependent conditions are likely be reduced on well resolved scales (e.g. the reduced impact of cross-lobe age variations) but, when integrated, the compound effect of small variations may be responsible for the observed discrepancy.

While such effects can plausibly account for the differences observed in active FR IIs, its applicability to both remnant sources and FR Is is less clear. In the case of 3C31, an intrinsically flatter than observed low frequency spectrum could potentially resolve the lack of change in the spectrum around the spectral break; however, if as is commonly assumed first order Fermi is the primary particle acceleration mechanism, this is restricted to values of $\alpha_{inj} \geq 0.5$ and so a significant discrepancy will still remain. Conversely, remnant galaxies appear to suffer from a flatter than expected low frequency spectrum and/or an unexpectedly sharp cutoff at higher frequencies. The limited number of remnant galaxies currently known means it is not yet possible to determine whether such effects are common to the remnant population and, if so, which is the primary cause of the discrepancy, but the results are indicative that additional considerations must be made within the models if they are to provide a robust description of these sources.

Given the frequency dependence of the CI models, the fact that they are unable to provide a reliable measure of the injection index and, even when model parameters are known a priori, provide either a poor description of a source's spectrum or significant discrepancies in the derived ages are observed, I suggest that these models do not provide a robust measure of a radio galaxy's spectral age. Revised integrated models of spectral ageing such as those currently being developed by Godfrey et al. (in prep) and Brienza et al. (in prep) which attempt to account for effects such as adiabatic expansion and non-uniform magnetic field strengths may go some way to resolving this issue, but it is likely that additional, time dependent factors such as those described above must also be incorporated before these models can be considered truly robust. I therefore recommend that the classical CI models should be used with caution when applied to radio galaxy populations in upcoming and future surveys.

\subsection{CI models as a method of source selection}
\label{selection}

\subsubsection{Remnant radio galaxies}
\label{remnantselection}

While the integrated models of spectral ageing appear to not provide a robust measure of spectral age this does not entirely rule out their usefulness in terms of candidate selection in large data volumes. As has been shown by \citet{brienza16} for the remnant radio galaxy BLOB1, and also appears to be the case for 3C28, the integrated low energy spectrum is much flatter than previously assumed meaning that, at least in some cases, remnant radio galaxy candidates cannot be selected based on an ultra-steep spectrum at low frequencies alone. Given that the upcoming surveys will be limited to a given frequency range (e.g. 110 - 280 MHz in LOTSS), there is therefore the need when searching for candidate remnant galaxies in surveys for complementary archival data to be used. This data is commonly in the form of low resolution catalogues (e.g. NVSS) or integrated flux values in the literature which, when combined with the large computational and human overheads that accompany resolved fitting, means that candidate searches currently \emph{must} be based on the integrated spectrum.

While I have shown that when model parameters are know a priori radio galaxies are often poorly described by integrated models, allowing the injection index to be a free parameter often provides a better fit to the overall spectrum. As it has already been established that any ages derived from these models are unlikely to be robust, allowing this parameter to vary may provide a method of selecting remnant radio galaxies in large data volumes due to the significant difference in the shape of the CI and CI off models.

Two main methods are currently used in the search for remnant galaxies: a steep ($\geq\,$ 1.2) spectral index at high frequencies, and a spectral curvature value ($\rm{SPC} = \alpha_{high} - \alpha_{low}$; \citealp{murgia11}) greater than 1.1. However, both are based on the classical assumptions made about the low frequency spectrum of radio galaxies, particularly in FR IIs where it is observed to be much steeper than expected. When using the classical assumption of $\alpha_{inj} \approx 0.6$ combined with Equation \ref{alphahigh} we see that for active sources the spectrum at high frequencies is expected to take values of around 1.1. Allowing for reasonable variations in the injection index, anything steeper than  $\alpha_{high} = 1.2$ should therefore be a strong candidate for a remnant galaxy but, in the case of FR IIs where $\alpha_{inj}$ is now known to take values of around 0.7 to 0.8, such a sample would include a large number of active sources. One solution would be to simply increase the cut off for $\alpha_{high}$ to steeper values but this would then exclude FR I remnants which exhibit a flatter injection index.

Selection based on SPC also encounter similar problems, with 3C28 proving a prime example of a source which would be missed by such a selection criteria. Considering the spectral index between 22 and 178 MHz ($\alpha_{low} = 1.08$) and between 1.4 and 7.9 GHZ ($\alpha_{high}=1.97$) results in a SPC value of only 0.9, well below the required selection criteria. While these methods have found some success in finding radio remnants, they are likely to be limited to only a subset of the population as a whole and may be a contributing factor to why so few FR II remnants have been found. CI model comparison may therefore provide an improved method of source selection over these classical methods.

For 3C28, BLOB1, and the recent study of the remnant B2 0924+30 by \citet{shulevski16}, the CI off model provides a \emph{significantly} better fit than the standard CI model due to the rapid steepening of the spectrum at higher frequencies. For example, using a flatter injection index of $0.75$ for 3C28 results in a reduced $\chi^{2}$ value of 1.85 for the CI off model compared to 248 for the standard CI model. While the model fit remains statistically rejected, the large difference between the two models provides a strong indication that the source is no longer active which can then be confirmed by follow up observations at higher resolutions or by alternative measures such as its morphology or lack of core emission.

Determining the number of extra remnant sources which could potentially be discovered using this method, how complete a sample this would allow, and what selection criteria should be used to reduce the effects of contamination from non-remnant sources requires further investigation using mock catalogues and/or well studied fields (e.g. the Lockman hole; \citealp{mahony16}) which is beyond the scope of this paper. However, the application of this comparative approach should afford some improvement on those currently in use and, on the timescales of the ongoing surveys, provides potentially the best method for identifying samples of remnant radio galaxies.

\subsubsection{High redshift radio galaxies}
\label{highzselection}

An additional possibility for the use of CI models in radio surveys is aiding in the selection of high redshift radio galaxies. The integrated spectral index of radio galaxies is known to become steeper with increasing redshift \citep{athreya98} and has therefore become a common method of source selection; however, this spectrum is also highly dependent on the age of the source and its current phase of activity (i.e. active or remnant). Crossmatching samples of radio galaxies to their optical counterparts in order to obtain robust redshift measurements, particularly for the most distant objects, requires a significant amount of observing time and so for large samples a clean initial sample is desirable.

While it is not possible to reliably determine the redshift of radio galaxies from the CI models alone, the $z - \alpha$ relation must at least in part be due to an increase in losses from the inverse-Compton scattering of the CMB. As this effect scales differently from the age of the source (Equation \ref{break}), by setting the redshift as a free parameter in a similar manner to the injection index discussed above, it is theoretically possible to determine which sources are likely to be high redshift and which are instead nearby old or remnant radio galaxies. Such a method is unlikely to ever provide as robust a measure of redshift compared to those determined spectroscopically but, for very large samples, may provide a method of increasing the efficiency of such searches by removing false positives.

Testing of this method against mock catalogues is again beyond the scope of this paper but note in order for this method to be reliable, the spectrum of each target would need to well constrained so that the additional free parameter space could be be fully explored. Such observations are at the current limit of telescope capabilities for all but the oldest/most distant radio galaxies where such a break occurs at or below, for example, VLA P-band frequencies. However, instruments such as MeerKAT will extend this range to the GHz regime and, looking forward to the SKA era where such data will be commonplace, provides a potentially feasible component of source selection for high redshift radio galaxies.

\section{Conclusions}
\label{conclusions}

In this paper, I have presented the first comparison between integrated models of spectral ageing and recent investigations using well resolved, modern techniques, focusing on implications of their usage in upcoming surveys and for the determination of the science possible with the next generation of radio telescope. The main findings are as follows:

\begin{itemize}
\item Integrated model fitting is unable to reliably recover key parameters such as the injection index.\\
\item When model parameters are known a priori, fitting of the integrated spectrum is highly frequency dependent and often provides a poor description of the source's spectrum.\\
\item Integrated models of spectral ageing are currently unable to provide robust ages of a source, with a disparity of up to a factor of 6 being observed compared to their well resolved counterparts.\\
\item Although inconsistencies exist between CI models and observations, they remain a potentially useful tool for selecting candidates in the search for remnant radio galaxies.
\end{itemize}

I therefore conclude that the application of standard CI (off) models to the integrated spectrum of radio sources should, in their current form, be treated with caution. However, they may still provide a useful tool for identifying remnant radio galaxy candidates in cases where a steep spectrum at long wavelengths is alone unable to determine the current phase of a radio galaxies life cycle, which can then be followed up by additional high resolution observations. I suggest that they may also provide a method of providing cleaner samples of high redshift radio galaxies when combined with selection based on the $z - \alpha$ relation.

\section*{Acknowledgements}
\label{acknowledgements}
I wish to thank Volker Heesen for providing the data for 3C31 and the anonymous referee for their useful comments and suggestions which have helped improve this paper. This research was partly funded by the European Research Council under the European Union's Seventh Framework Programme (FP/2007-2013)/ERC Advanced Grant RADIOLIFE-320745. This research has made use of the NASA/IPAC Extragalactic Database (NED), which is operated by the Jet Propulsion Laboratory, California Institute of Technology, under contract with the National Aeronautics and Space Administration. 

%\bibliographystyle{mn2e}
%
%\def\newblock{\hskip .11em plus .33em minus .07em}
%\bibliography{civsresolved}

\bibliographystyle{mnras}
\bibliography{civsresolved}

% Don't change these lines
\bsp	% typesetting comment
\label{lastpage}
\end{document}